\documentclass[]{emulateapj}

\shorttitle{The distribution of SiO in the CSE around IRC\,+10216}
\shortauthors{Sch\"oier et al.}

\begin{document}


\title{The distribution of SiO in the circumstellar envelope around IRC\,+10216}


\author{Fredrik L. Sch\"oier}
\affil{Stockholm Observatory, AlbaNova University Center, SE-106 91 Stockholm, Sweden}
\email{fredrik@astro.su.se}

\author{David Fong}
\affil{Harvard-Smithsonian Center for Astrophysics,  60 Garden Street, Cambridge, MA 02138, USA}

\author{Hans Olofsson}
\affil{Stockholm Observatory, AlbaNova University Center, SE-106 91 Stockholm, Sweden}
\affil{Onsala Space Observatory, SE-439 92 Onsala, Sweden}

\author{Qizhou Zhang}
\affil{Harvard-Smithsonian Center for Astrophysics, 60 Garden Street, Cambridge, MA 02138, USA}

\author{Nimesh Patel}
\affil{Harvard-Smithsonian Center for Astrophysics, 60 Garden Street, Cambridge, MA 02138, USA}



\begin{abstract}
New interferometric observations of SiO $J$\,$=$\,5\,$\rightarrow$\,4 circumstellar line emission around the carbon star {IRC+10216}, using the Submillimeter Array, are presented. Complemented by multi-transition single-dish observations, including infrared observations of ro-vibrational transitions, 
detailed radiative transfer modelling suggests that the fractional abundance of SiO in the inner part of the envelope, between 
$\approx$\,3--8 stellar radii, is as high as $\approx$\,1.5\,$\times$\,10$^{-6}$.
This is more than an order of magnitude higher than predicted by equilibrium stellar atmosphere chemistry in a carbon-rich environment and indicative of the importance of non-LTE chemical processes. 
In addition to the compact component, a spatially more extended ($r_{\mathrm{e}}$\,$\approx$\,2.4\,$\times$\,10$^{16}$\,cm) low-fractional-abundance ($f_0$\,$\approx$\,1.7\,$\times$\,10$^{-7}$) region is required to fit the observations. This suggests that the majority of the SiO molecules are effectively accreted onto dust grains in the inner wind while the remaining gas-phase molecules are eventually photodissociated at larger distances.
Evidence of departure from a smooth wind is found in the observed visibilities, indicative of density variations of a factor 2 to 5 on an angular scale corresponding to a time scale of about 200 years.
Additionally, constraints on the velocity structure of the wind are obtained.

\end{abstract}


\keywords{stars: abundances -- stars: AGB and post-AGB -- stars: carbon -- stars: circumstellar matter -- stars: individual (IRC\,+10216) -- stars: mass loss}



\section{Introduction}
IRC\,+10216 (CW Leo) is a cool red giant star located on the asymptotic giant branch (AGB). It is classified as a carbon star \citep{Miller70, Herbig70} with a C/O-ratio of $\gtrsim$\,1.4 in its atmosphere  \citep{Winters94,Glassgold96} and is currently losing mass at a rate of $\approx$\,1--2\,$\times$\,10$^{-5}$\,M$_{\odot}$\,yr$^{-1}$, as determined from  a wide variety of observations from infrared to radio wavelengths and using molecular line as well as  dust continuum emission \citep[e.g.,][]{Keady88,Groenewegen98, Schoeier01, Schoeier02b}. It is commonly thought that {IRC\,+10216} is in its last evolutionary stage on the AGB before the planetary nebula ejection phase. The high mass-loss rate of {IRC\,+10216} makes direct observations of the stellar atmosphere difficult, and the future evolution of the star is possibly best followed through monitoring of its circumstellar material, although some stellar light emerges due to the in-homogeneous nature of the wind. Departures from a homogeneous wind are evident from images of interstellar light scattered in the circumstellar envelope (CSE) around  {IRC\,+10216} \citep{Mauron99, Mauron00}. The high mass-loss rate of {IRC\,+10216}, coupled with its close proximity to the Sun ($\approx$\,120\,pc), has made it the prime target for studies of not only carbon stars, but also AGB stars in general. 

   \begin{figure*}
    \label{sma_fig}
   \centering{   
   \includegraphics[width=15cm,angle=-90]{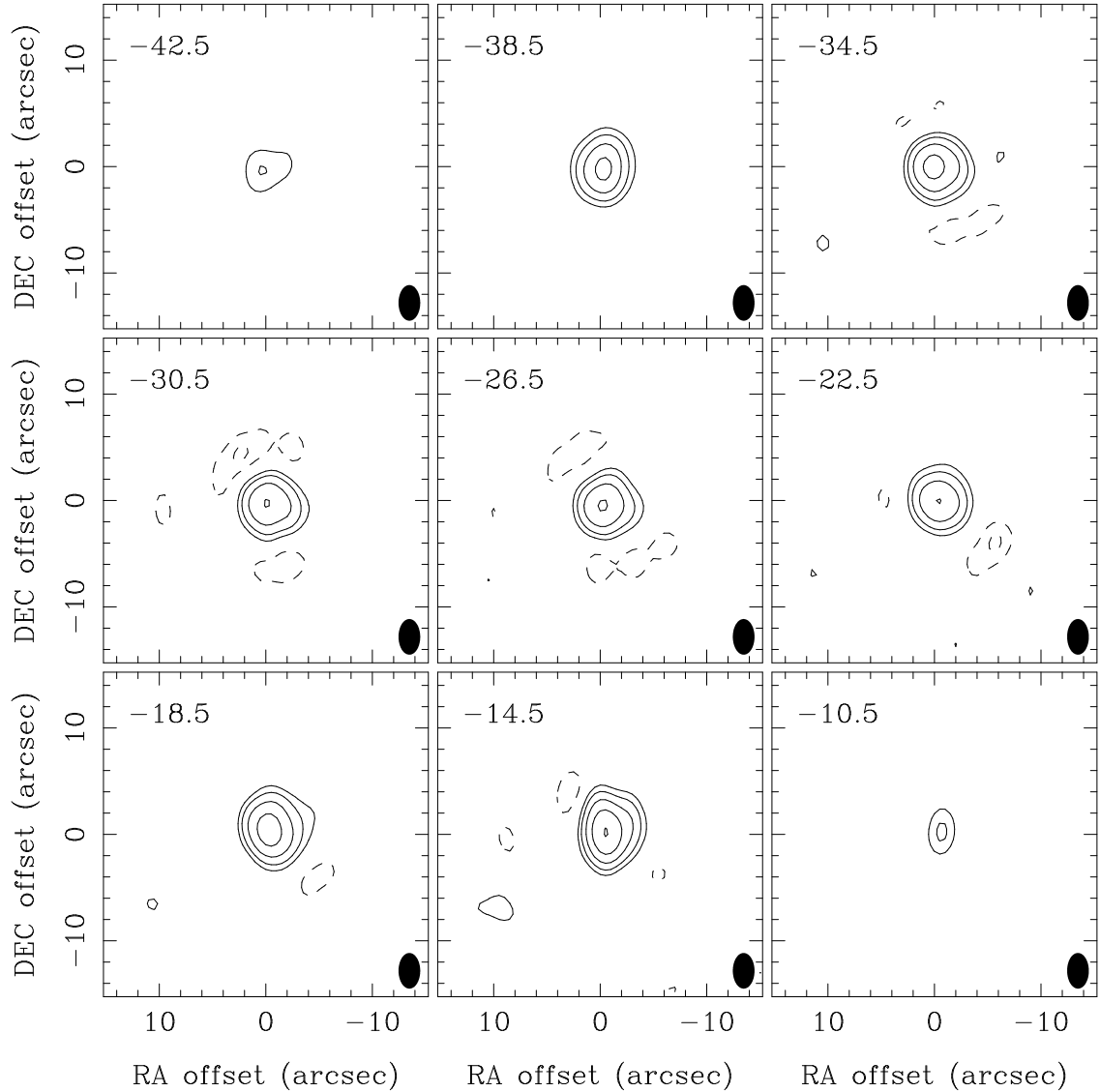}
   \caption{Velocity channel maps of SiO $J$\,$=$\,5\,$\rightarrow$\,4 line emission from IRC\,+10216 obtained using the SMA.  
The contour levels are $1.0n$ Jy~beam$^{-1}$, where $n=-4,-2,2,4,8,16,32$ (negative values have dashed contours), and the beam size
is $3\farcs3\times2\farcs0$ with a position angle of $0\fdg2$ as indicated in the lower right corner of each panel.  The velocity channels (given in the
LSR frame and indicated in the upper left corner) have been binned to 4\,km\,s$^{-1}$. The systemic velocity is $-$26\,km\,s$^{-1}$ as determined from CO observations.
Offsets in position are relative to $\alpha_{2000} = 09^{\rm h} 47^{\rm m} 57\fs39$, $\delta_{2000} = 13\degr 16\arcmin 43\farcs9$.}}
   \end{figure*}

Presently, there are 63 molecular species detected in AGB stars with about half of them detected in only {IRC\,+10216}  \citep{Olofsson06}. A significant number of these molecules are unique to the circumstellar medium which makes the study of stellar winds important also for testing our knowledge of interstellar chemistry in general. Most of the abundance estimates are based on rather simple methods and are typically order of magnitude estimates. Detailed chemical modelling of  {IRC\,+10216} does a reasonable job in explaining many of the observed abundances \citep{Millar03}. However, there are some notable exceptions. The detection of H$_2$O towards {IRC\,+10216}  by the {\em SWAS} satellite \citep{Melnick01}, and reconfirmed by {\em Odin} observations \citep{Hasegawa06}, came as a surprise.
Recently, \citet{Hasegawa06} derived an abundance of NH$_3$, based on observations by the Odin satellite, that is four orders of magnitude higher than predicted by stellar equilibrium chemistry \citep{Cherchneff92}, and processing by shocks does not seem to increase the abundance any further \citep{Willacy98}. A possible explanation  for these high abundances could be Fischer-Tropsch catalytic processes as suggested by \citet{Willacy04} or, in the case of H$_2$O, evaporation of Kuiper-belt like objects \citep{Melnick01}.

   \begin{figure*}
   \centering{   
   \includegraphics[width=17cm]{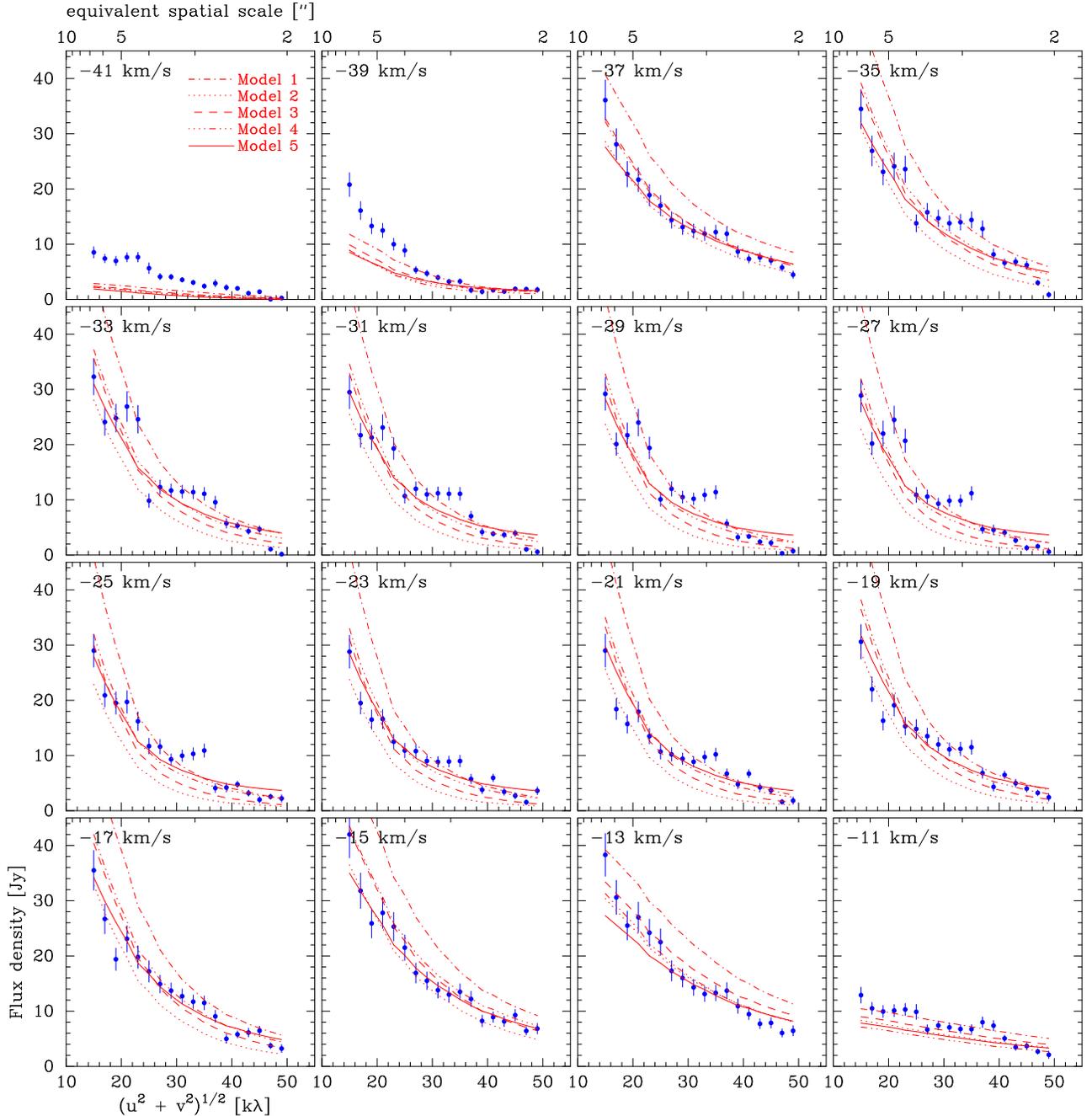}
   \caption{Azimuthally averaged visibility amplitudes (mean value taken over 2\,km\,s$^{-1}$ bins) obtained towards IRC\,+10216 using the SMA, as a function of distance to the phase centre in k$\lambda$ (also shown on the upper abscissa is the equivalent spatial resolution in $\arcsec$). The observations are compared with various models for the SiO abundance distru'bution. Models 1--4 contain a single SiO abundance component whereas Model 5 in addition has a compact high-abundance component (see text for details).}
    \label{sma_uvall}}
   \end{figure*}

The first two more detailed studies of circumstellar abundances in larger samples of sources have been performed by \citet{Delgado03b} and \citet{Schoeier06a} for SiO in 45 M-type (C/O\,$<$\,1 in the photosphere) AGB stars and 19 carbon stars (C/O\,$>$\,1 in the photosphere), respectively. Average SiO fractional abundances were obtained from a detailed radiative transfer analysis of multi-transition single-dish observations. 
Interestingly, for the M-type AGB stars the 
derived abundances are generally much lower than expected from photospheric equilibrium chemistry ($\approx$\,5\,$\times$\,10$^{-5}$, Duari et al.\ 1999\nocite{Duari99}). For the carbon stars, on the other hand, the derived abundances are on the average two orders of magnitude higher than predicted by photospheric equilibrium chemistry ($\approx$\,5\,$\times$\,10$^{-8}$, Millar 2003\nocite{Millar03}). 
In fact, when comparing the two distributions of SiO fractional abundances there appears to be no way of distinguishing a C-rich chemistry from that of an O-rich based on an estimate of the circumstellar SiO abundance alone. Moreover,  there is a clear trend that the SiO fractional abundance decreases as the mass-loss rate of the star increases, as would be the case if SiO is accreted onto dust grains. 

Further support for such a scenario comes from interferometric observations of the two M-type  AGB stars {R~Dor} and {L$^2$~Pup} performed by \citet{Schoeier04b}. In their analysis they found evidence of an inner compact component of high fractional abundance, consistent with predictions from stellar atmosphere chemistry. In addition, an extended  low-abundance  component, as expected if SiO is effectively depleted onto grains in the inner wind, was required in order to fit the observations.

Presented here are new interferometric observations of SiO $v$\,=\,0, $J$\,$=$\,5\,$\rightarrow$\,4 line emission from the circumstellar envelope around the carbon star IRC\,+10216 obtained by the Submillimeter Array (SMA). Supplemented by additional multi-transition single-dish radio and infrared observations the radial abundance profile of SiO is determined. In addition, information on the acceleration region close to the photosphere is obtained. 

%
%

\section{Observations}





\label{sect_obs}
SMA\footnote{The Submillimeter Array is a joint project between the Smithsonian
Astrophysical Observatory and the Academia Sinica Institute of Astronomy and
Astrophysics, and is funded by the Smithsonian Institution and the Academia
Sinica.} observations of the SiO $J$\,$=$\,5\,$\rightarrow$\,4 line at 217.10498 GHz
were made on 2005 February 23 using the compact array configuration with seven
6\,m antennas.  Details of the SMA are described by \citet{Ho04}.  The
weather was excellent, with a 225\,GHz atmospheric opacity of 0.05 (measured
at the nearby Caltech Submillimeter Observatory) and system temperatures (DSB)
ranging between 150 to 300\,K.  The projected baselines ranged from
$11-50$\,k$\lambda$ resulting in a synthesized beam 
of $3\farcs3\times2\farcs0$, with a PA of 0$\fdg$2 using uniform weighting. 
The digital
correlator has a bandwidth of 2\,GHz and the spectral resolution was 0.8125\,MHz,
corresponding to a velocity resolution of 1.1\,km\,s$^{-1}$.  Phase and amplitude
calibration were performed on the quasars 0851+202 and 1058+015.
Bandpass calibration was obtained on Callisto and Jupiter.  Observations of
Callisto provided the flux calibration; the uncertainty of the flux scale is
estimated to be $\sim$20\%.  The data were calibrated using the MIR software
package developed originally for the Owens Valley Radio Observatory and adapted
for the SMA.  The calibrated visibility data were imaged and CLEANed using MIRIAD.

   \begin{figure}
   \centering{   
   \includegraphics[width=7cm,angle=-90]{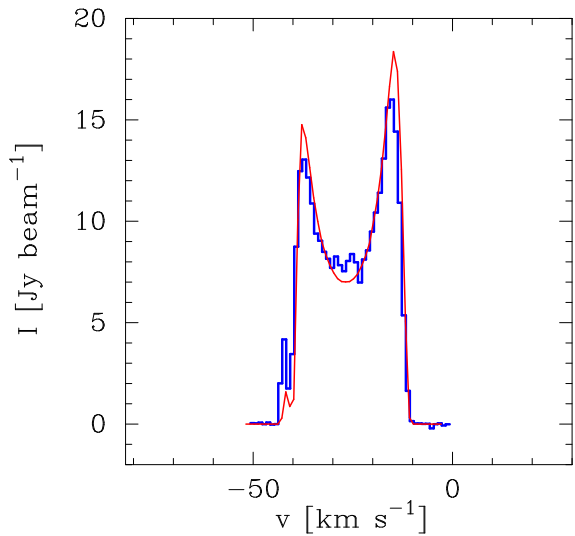}
   \caption{Spectrum at the central pixel in the CLEANed map (histogram) overlayed by the best-fit model (Model 5; solid line). The velocity resolution is 1\,km\,s$^{-1}$.}
    \label{sma_spectrum}}
   \end{figure}

The actual analysis and comparison with models are carried out
in the $uv$-plane to maximize the sensitivity and resolution of the
data.  Thus, we expect to obtain usable information on scales as low as
$2\arcsec$, corresponding to the longest baselines.


%
\begin{figure}
\centerline{\includegraphics[width=7cm,angle=-90]{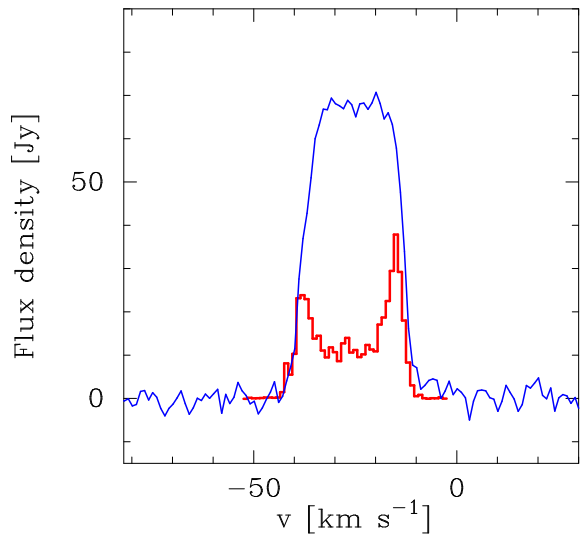}}
  \caption{SiO $v=0, J=5\rightarrow4$ spectra for  IRC\,+10216. 
 The solid line is the SEST single-dish observation \citep{Delgado03b}, whereas the histogram shows the spectrum at the phase centre  derived from the SMA data 
 convolved to the size of the SEST beam of $23\arcsec$. The velocity resolution is 1\,km\,s$^{-1}$.}
  \label{sma_recover}
\end{figure}
   \begin{figure}
   \centering{   
   \includegraphics[width=8.5cm]{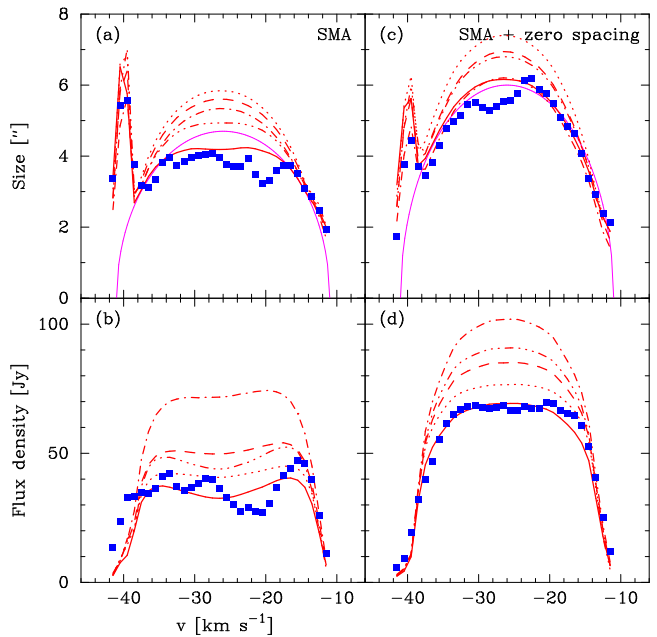}
   \caption{Size ({\em{FWHM}}; panels (a) and (c)) and flux (panels (b) and (d)) of the observed SiO $J$\,$=$\,5\,$\rightarrow$\,4 line emission toward  
 IRC\,+10216  as function of velocity channel (given in LSR frame), averaged in 1 km\,s$^{-1}$ bins, as estimated from Gaussian fits to the visibilities.  Panels (a) and (b) are fits to the SMA data alone while panels (c) and (d) have the flux fixed at the zero-spacing value.
 The observations are compared with the results from single SiO component models with (Model 4; dashed-dot-dot-dot line) and without (Models 1-3; dash-dot, dotted and  dashed-line, respectively) a radial velocity gradient and a model with an additional compact high abundance component (Model 5; solid line). A fit to the data using Eq.~\ref{sizevel} for the size of the emitting region is also shown (panels (a) and (c); thin solid line).}
    \label{sma_size}}
   \end{figure}
%

%
%

\section{SiO brightness distribution}
The velocity channel maps of the SiO $J$\,$=$\,5\,$\rightarrow$\,4  line emission obtained by the SMA, presented in Fig.~\ref{sma_fig}, show that the brightness distribution towards IRC\,+10216 appears to have an overall circular symmetry and suggests that the emission 
 is moderately resolved. At this level of resolution no signs of deviations from a homogeneous wind such as mass loss modulations or clumps are evident. 
 
In Fig.~\ref{sma_uvall} the azimuthally-averaged visibility amplitudes, mean value taken over 2\,km\,s$^{-1}$ bins, are plotted as a function of distance to the phase centre in k$\lambda$. It is evident that there is some degree of patchiness in the distribution. 
The visibility amplitudes show two regions where the amplitudes flatten out: at around 20 and 32\,k$\lambda$.  The plateau at 20\,k$\lambda$ is most evident at the blue-shifted velocities, while the plateau at 32\,k$\lambda$ is seen at both blue- and red-shifted velocities.  At the extreme velocities, these plateau regions are not apparent. The fact that these patterns do not appear constant from velocity channel to velocity channel make us believe that they represent real structure in the SiO brightness distribution.
The spatial scales involved are $\approx$\,5$\arcsec$ and $\approx$\,3$\arcsec$ for the 20 and  32\,k$\lambda$ plateau regions, respectively. From the radiative transfer modelling performed in Sect.~\ref{sma} the spread in the observed visibilities can be accounted for by SiO abundance modulations of a factor $\approx$\,2--5, or a similar change of the H$_2$ density. Both the time scales, $\approx$\,200\,yr, and density modulations are consistent with those obtained from the scattered light images of \citet{Mauron99,Mauron00}.
 
 The spectrum at the central pixel in the CLEANed map  is presented in Fig.~\ref{sma_spectrum} (histogram). The spectrum is characteristic of a well-resolved, with respect to the beam, emitting region of moderate optical depth. Optical depth effects are apparent on the blue-part of the spectrum in Fig.~\ref{sma_spectrum} which generally has a lower intensity  than the red part due to self-absorption of  emission  in gas moving towards the observer.
 
As interferometers lack sensitivity to large scale emission  it is of interest to ascertain the missing flux of the SMA data. In Fig.~\ref{sma_recover} the SMA
SiO $J$\,$=$\,5\,$\rightarrow$\,4  data has been convolved with a $23\arcsec$ circular beam (histogram) to represent the beam of the Swedish-ESO Submillimetre Telescope (SEST). 
The SEST spectrum has been converted from main beam brightness temperature scale to Jy using 
\begin{equation}
\label{K2Jy}
S = \eta_{\rm mb}  \Gamma^{-1} T_{\mathrm{mb}}, 
\end{equation}
where the main-beam efficiency $\eta_{\rm mb}$\,=\,0.5 
and the sensitivity $\Gamma^{-1}$\,=\,40\,Jy\,K$^{-1}$
The telescope parameters are taken from the SEST homepage\footnote{\tt{www.ls.eso.org/lasilla/Telescopes/SEST/SEST.html}}.
A comparison with the equivalent SEST spectrum from 
\citet{Schoeier06a} (solid line) shows that much of the emission is resolved out by the interferometer. 
In addition, the CLEANed image of an extended source with missing short-spacing
data results in an extended depression of negative surface brightness
on which the source emission resides.  This artifact is most evident
in the central velocity channels in Fig.~\ref{sma_fig}, where the real envelope
emission is expected to be the most extended.  The negative
feature will further reduce the size and flux of the emission in those channels. In the central velocity channels of Fig.~\ref{sma_recover} only about 10\,Jy is recovered after the cleaning compared to $\approx$\,29\,Jy that the interferometer actually measures (Fig.~\ref{sma_uvall}).
  
Assuming that the emission has an overall spherical symmetry, circularly symmetric Gaussians have been fitted to the visibilities in 1\,km\,s$^{-1}$ velocity bins. The resulting {\em FWHM}s and fluxes of the circular Gaussians fitted to the visibilities are plotted in Figs.~\ref{sma_size}a and ~\ref{sma_size}b.
There appears to be no systematic trend in the derived offsets with velocity.  The flux in each velocity bin is significantly higher than the flux obtained from the full map (Fig.~\ref{sma_recover}; histogram). The estimated flux from the Gaussian fit is, however, the predicted zero-spacing flux based on the observed visibilities. Hence, it tries to compensate for the missing flux in the observations. The fluxes reported in Fig.~\ref{sma_size}b are lower than the total flux picked up by the SEST single-dish observation 
(Fig.~\ref{sma_recover}; solid line) indicating a departure of the brightness distribution from that of a  Gaussian at larger spatial scales.

The overall variation in size with line-of-sight velocity (Fig.~\ref{sma_size}a) is not as expected for a well resolved envelope expanding at a constant velocity, where gas moving orthogonal to the line of sight subtends a larger solid angle than radially moving gas near the extreme velocities, as described by
\begin{equation}
\label{sizevel}
R(v) = R_{\mathrm s} \left [ 1- \left ( \frac{v-v_{\mathrm{lsr}}}{v_\mathrm{e}} \right ) ^2 \right ] ^{1/2},
\end{equation}
where $R(v)$ is the estimated size at velocity $v$, $R_{\mathrm s}$ the total size of the emitting region, $v_{\mathrm{lsr}}$ the systemic velocity (LSR), and $v_{\mathrm{e}}$ the expansion velocity. Using Eq.~\ref{sizevel}, with $R_{\mathrm s}$\,=\,4\farcs7, $v_{\mathrm{lsr}}$\,=\,$-26.5$\,km\,s$^{-1}$ and $v_{\mathrm{e}}$\,=\,15\,km\,s$^{-1}$, does not provide an acceptable fit (Fig.~\ref{sma_size}a; thin solid line).
The largest angular extent instead appears close to the blue part of the emission with $5\farcs4\pm0\farcs1$ in the velocity range $-40$ to $-38$ km~s$^{-1}$ (LSR). This corresponds to a radial size of $1.0\times 10^{16}$ cm at a distance of 120 pc (see below). It should be noted that in this region the line profile is heavily affected by self-absorption as seen in Fig.~\ref{sma_spectrum}. 

Self-absorption would naturally explain the larger spatial extent near the blue edge of the spectrum since photons emitted by the gas close to the star, and which in the optically thin case would have no further interaction with the gas, can become absorbed and re-emitted at larger distances from the star thereby increasing the size of the brightness distribution. The lower intensity of the emission naturally follows from the decrease of density and temperature with radial distance from the star.

However, there is some concern that the size-velocity relation in Fig.~\ref{sma_size}a is affected by the lack of short-spacing information. The predicted zero-spacing flux near the systemic velocity from the Gaussian fits (Fig.~\ref{sma_size}b) is about a factor of two lower than what is observed using the SEST (Fig.~\ref{sma_recover}; solid line). If instead the flux of the Gaussian fit is fixed at the SEST value (Fig.~\ref{sma_size}d), a different size-velocity relation is found, as illustrated in Fig.~\ref{sma_size}c. The observed size-velocity relation is now better described by Eq.~\ref{sizevel}, in this case with  $R_{\mathrm s}$\,=\,6\farcs0 (thin soild line in Fig.~\ref{sma_size}c), corresponding to 1.1\,$\times$\,10$^{16}$\,cm$^{-1}$ at the distance of IRC+\,10216. It is noted that the size estimate near the blue edge of the spectrum is still significantly larger than expected from Eq.~\ref{sizevel} due to the self-absorption.



%
   \begin{figure}
   \centering{   
   \includegraphics[width=7cm,angle=-90]{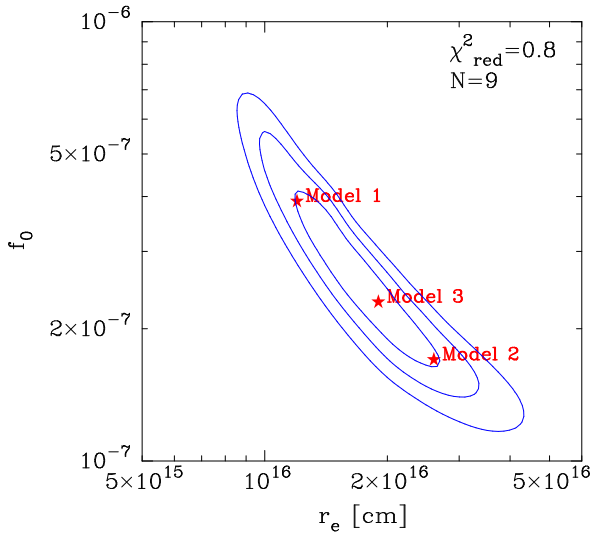}
   \caption{$\chi^2$ map from the single dish-modelling of radio-line data performed by  \citet{Schoeier06a} showing the sensitivity of the excitation analysis to the adopted SiO fractional abundance ($f_0$) and envelope size ($r_{\mathrm e}$). Contours are drawn at the 1, 2, and 3$\sigma$ levels. Also indicated is the lowest reduced $\chi^2$ ($\chi^2_{\mathrm{red}}=\chi^2_{\mathrm{tot}}/(N-2)$) in the map.}
   \label{chi2}}
   \end{figure}
%

%
%

\section{SiO excitation analysis}
\label{sect_model}
\subsection{Radiative transfer model}
In order to determine the molecular excitation in the CSE a detailed non-LTE radiative transfer code, based on the Monte Carlo method,
was used.  The code is described in detail in \citet{Schoeier01} and has been
benchmarked, to high accuracy, against a wide variety of molecular-line radiative
transfer codes in \citet{Zadelhoff02} and van~der~Tak et al.\ (in prep.).  
The CSE around {IRC+10216} is assumed to be spherically symmetric, produced by a constant mass-loss rate ($\dot{M}$), and to expand at a constant velocity ($v_{\mathrm e}$). A velocity gradient (acceleration region) is also tested in Sect.~\ref{sect_vel}.

The physical properties, such as the density, temperature, and kinematic structure prevailing in the circumstellar envelope around {IRC+10216} have been determined in \citet{Schoeier00,Schoeier01}, \citet{Schoeier02b}, and \citet{Schoeier06a},   based on radiative transfer modelling of multi-transition CO line observations, from millimetre to IR wavelengths. This model, where the circumstellar envelope is formed by a mass-loss rate of 1.5\,$\times$\,10$^{-5}$\,M$_{\odot}$\,yr$^{-1}$ and expanding at a velocity of 14.0\,km\,s$^{-1}$, is used as input to the SiO excitation analysis.

The excitation analysis includes radiative excitation through the first vibrationally excited ($v$\,=\,1) state, for SiO at 8\,$\mu$m.
\citet{Schoeier06a}  found that for SiO it is particularly important that this is treated correctly. Relevant molecular data are summarized in \citet{Schoeier05a} and are made publicly available through the {\em Leiden Atomic and Molecular Database} (LAMDA){\footnote{\tt http://www.strw.leidenuniv.nl/$\sim$moldata}}. For {IRC+10216}  thermal dust emission provides the main source of infrared photons which excite the $v$\,=\,1 state. The addition of a dust component in the Monte Carlo scheme is straightforward as described in \citet{Schoeier02b}. The dust-temperature structure and dust-density profile for {IRC+10216} are obtained from 
radiative transfer modelling of the spectral energy distribution using {\em Dusty} \citep{Ivezic97}. Details on the dust modelling can be found in  \citet{Schoeier06a}. The best-fit model is obtained for an optical depth at 10\,$\mu$m of 0.9, a dust-condensation temperature of 1200\,K ($r_0$\,=\,1.7\,$\times$\,10$^{14}$\,cm) and an effective stellar temperature of 2000\,K. The luminosity is 9600\,L$_{\odot}$, obtained from a period-luminosity relation \citep{Groenewegen96}, and the corresponding distance is 120\,pc. It should be noted that  this period-lumonosity relation was statistically derived for 54 carbon stars with pulsational periods in the range 150\,$-$\,520 days. An additional uncertainty comes from the fact that the period of {IRC+10216}  is 630 days.

\subsection{The SiO abundance distribution}
\label{sma}
The abundance distribution of SiO is initially assumed to be described by a Gaussian
\begin{equation}
\label{eq_distr}
f(r) = f_0\, \exp \left(-\left(\frac{r}{r_{\mathrm e}}\right)^2 \right),
\end{equation}
where $f$\,$=$\,$n\mathrm{(SiO)}/n\mathrm{(H_2)}$, i.e., the ratio of the number density of SiO molecules to that of H$_2$ molecules. Here, $f_0$ denotes the photospheric fractional abundance of SiO and it is assumed that some process effectively destroys the molecules at $r$\,$>$\,$\,r_{\mathrm e}$ such as, e.g.,  photodissociation by the ambient interstellar uv-field. \citet{Schoeier06a} found that such an abundance distribution could explain multi-transition single-dish radio-line SiO observations of IRC\,+10216 well, with $f_0$\,$=$\,$2.8$\,$\pm$\,1.1\,$\times$\,10$^{-7}$ and $r_{\mathrm e}$\,$=$\,$1.9$\,$\pm$\,0.7\,$\times$\,10$^{16}$\,cm (see Fig.~\ref{chi2}). 

Fig.~\ref{sma_uvall} shows the result of applying the ($u,v$) sampling of the SMA observations to various SiO envelope models. 
All models (1--3) are within the 1\,$\sigma$ confidence level when compared to the constraints put by the single-dish data (see Fig.~\ref{chi2}). A model with $f_0$\,$=$\,3.9\,$\times$\,10$^{-7}$ and $r_{\mathrm{e}}$\,$=$\,1.2\,$\times$\,10$^{16}$\,cm (Model 1; dash-dotted line) provides a poor fit  with too much flux on most baselines while a model with $f_0$\,$=$\,1.7\,$\times$\,10$^{-7}$ and $r_{\mathrm{e}}$\,$=$\,2.6\,$\times$\,10$^{16}$\,cm (Model 2; dotted line) generally has too little flux and gives also a poor fit. However, a model with $f_0$\,$=$\,2.3\,$\times$\,10$^{-7}$ and $r_{\mathrm{e}}$\,$=$\,1.9\,$\times$\,10$^{16}$\,cm (Model 3; dashed line) provides a better overall fit, illustrating the usefulness of the SMA data in further constraining the SiO abundance distribution. The models are summarized in Table~\ref{models}. 

The models are particularly poor at reproducing the observed visibilities in the most blue-shifted part of the spectrum, between $-42$\,km\,s$^{-1}$ to $-38$\,km\,s$^{-1}$. This indicates  that there is too much self-absorption of the emission in the models. This could possibly indicate that the medium is clumpy to some degree. Nevertheless, the size-velocity behaviour near the blue-shifted edge as seen in Fig.~\ref{sma_size} 
is nicely reproduced by the model and the large observed size is explained by self-absorption in the gas moving towards the observer. 

Models with a constant expansion velocity, such as models 1--3, provide poor fits to the observed size-velocity distribution closer to the systemic velocity. The effect of a velocity field in explaining the observed size-velocity distribution and departures from the single Gaussian fractional abundance distribution of SiO are tested in Sect.~\ref{sect_vel}, where also additional constraints provided by ro-vibrational transitions in the infrared are used. 

%
%

%
   \begin{figure}
   \centering{   
   \includegraphics[width=8.5cm]{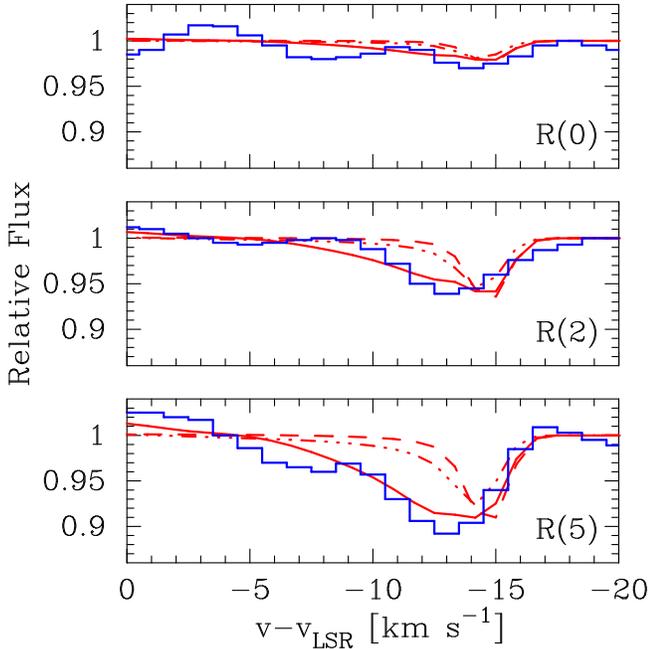}
   \caption{Calculated SiO  $v$\,$=$\,1\,$\rightarrow$\,0 $J$\,$=$\,1\,$\rightarrow$\,0 $R(0)$, $v$\,$=$\,1\,$\rightarrow$\,0 $J$\,$=$\,3\,$\rightarrow$\,2 $R(2)$ and $v$\,$=$\,1\,$\rightarrow$\,0 $J$\,$=$\,6\,$\rightarrow$\,5 $R(5)$ line profiles for the best-fit single-component model  using $f_0$\,$=$\,2.3\,$\times$\,10$^{-7}$ and $r_{\mathrm{e}}$\,$=$\,1.9\,$\times$\,10$^{16}$\,cm, with (Model 4; dash-dot-dot-dot line) and without (Model 3; dashed line) a gradient in the expansion velocity of the wind. The solid line is a two-component model (Model 5) including a compact ($r_{\mathrm{e}}$\,$=$\,4.5\,$\times$\,10$^{14}$\,cm) high-fractional-abundance ($f_{\mathrm J}$\,$=$\,1.5\,$\times$\,10$^{-6}$) SiO region in addition to the more extended ($r_{\mathrm{e}}$\,$=$\,2.4\,$\times$\,10$^{16}$\,cm) low-fractional-abundance ($f_0$\,$=$\,1.7\,$\times$\,10$^{-7}$) region and an acceleration region. The histograms are the observations performed by  \citet{Keady93}.}
   \label{rovib}}
   \end{figure}
\begin{table}
\caption{Model summary$^{\mathrm a}$.}
\label{models}
$$
\begin{array}{lcccccccc}
\hline
\noalign{\smallskip}
& &
\multicolumn{1}{c}{f_{\mathrm J}} &  &
\multicolumn{1}{c}{f_0}  &&
\multicolumn{1}{c}{r_{\mathrm e}}  && 
\multicolumn{1}{c}{\beta} \\ 
& & 
&& 
&&
\multicolumn{1}{c}{[\mathrm{cm}]} 
&& 
\\
\noalign{\smallskip}
\hline
\noalign{\smallskip}
\mathrm{Model\ 1}  &&  \cdots  &&  3.9\times10^{-7}    && 1.2\times10^{16} && 0.0\\
\mathrm{Model\ 2}  &&  \cdots  &&  1.7\times10^{-7}    && 2.6\times10^{16} && 0.0 \\
\mathrm{Model\ 3}  &&  \cdots  &&  2.3\times10^{-7}    && 1.9\times10^{16} &&0.0 \\
\mathrm{Model\ 4}  && \cdots &&  2.3\times10^{-7}    && 1.9\times10^{16} && 0.5  \\
\mathrm{Model\ 5}  &&  1.5\times10^{-6}   &&  1.7\times10^{-7}    && 2.4\times10^{16} && 0.15 \\
\noalign{\smallskip}
\hline
\end{array}
$$
$^{\mathrm a}$ The abundance distribution of SiO is assumed to be a Gaussian described by $f_0$ and $r_{\mathrm{e}}$ (see Eq.~\ref{eq_distr}). $f_{\mathrm{J}}$ is an increased abundance in the inner region of the wind and $\beta$ the slope of the radial velocity gradient (see\ text\ for\ details).
\end{table}
   \begin{figure}
   \centering{   
   \includegraphics[width=7cm,angle=-90]{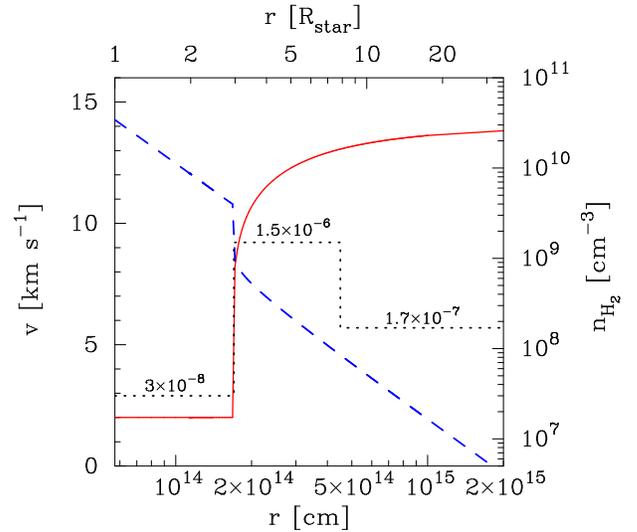}
   \caption{Adopted velocity field (solid line), based on a dynamical model, and the resulting H$_2$ density structure (dashed line) using a mass-loss rate of 1.5\,$\times$\,10$^{-5}$\,M$_{\odot}$\,yr$^{-1}$. Also shown is the derived SiO fractional abundance distribution (dotted line). }
    \label{structure}}
   \end{figure}
%

%


\subsection{Probing the inner wind}
\label{sect_vel}
Infrared observations of ro-vibrational transitions have been shown to be a sensitive tool for probing both the physical properties of the inner wind as well as the abundance structure \citep{Keady93, Monnier00}. However, the only carbon star to have been studied is {IRC\,+10216}.
Observations of SiO ro-vib transitions ($v$\,=\,1$\rightarrow$\,0) for {IRC\,+10216} have been performed by \citet{Keady93}. They found that the observed transitions (from $R$(0) up to $R$(5)), exhibiting classical P~Cygni-type line profiles, can be reasonably well modelled with an SiO fractional abundance of 8\,$\times$\,10$^{-7}$. Our best-fit single-dish model for IRC\,+10216, using $f_0$\,$=$\,2.6\,$\times$\,10$^{-7}$ and $r_{\mathrm{e}}$\,$=$\,1.7\,$\times$\,10$^{16}$\,cm, provides too little flux in these observed ro-vibrational transitions as illustrated in Fig.~\ref{rovib} for the  $R(0)$, $R(2)$ and $R(5)$ transitions (dashed lines; the original data from \citet{Keady93} are represented by the histograms). In particular the absorption feature produced in the model appears too narrow and is located at a too large velocity. A way to remedy this and fit the data better is to introduce a velocity field, i.e., an acceleration region in the inner envelope. Evidence for an acceleration region also comes from the estimated sizes from Gaussian fits in the $uv$-plane (Fig.~\ref{sma_size}), where the models ($1-3$), which behave as  expected for a resolved wind expanding at a constant velocity, give a poor fit to the size-velocity relation. The observations suggest that parts of the emission arises in the acceleration region. 

   \begin{figure*}
   \centering{   
   \includegraphics[width=17cm]{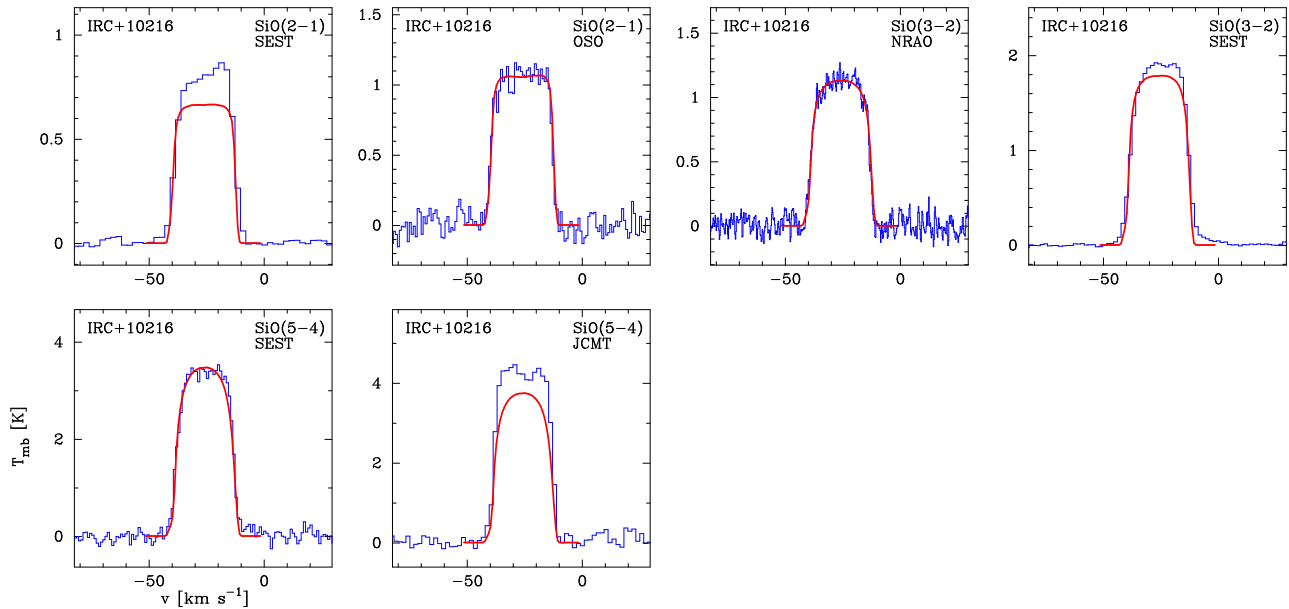}
   \caption{Best-fit two-component model (Model 5; solid lines) overlayed on observed single-dish spectra (histograms). This model also provides the best fit to the SMA and ro-vibrational data.}
   \label{single}}
   \end{figure*}

We find that by adopting a velocity field described by \citep{Habing94}
\begin{equation}
\label{velfield}
v(r) = v(\infty) \left(1-\frac{r_0}{r} \right)^{\beta},
\end{equation}
with $\beta$\,=\,0.5, it is possible to further  improve the fit to the observed line shape. Eq.~\ref{velfield} with $\beta$\,=\,0.5  is valid if the ratio of the dust and gas velocities is constant throughout the wind. When introducing a velocity gradient the density structure will also change according to the continuity equation, keeping the mass-loss rate constant. This has been taken into account in the modelling. All models are required to provide fits also to the single-dish radio-data within the 1$\sigma$ level.
In Fig.~\ref{rovib}  a model with $r_0$\,=\,1.7\,$\times$\,10$^{14}$\,cm (adopted to coincide with the dust condensation radius, see \citet{Schoeier06a}) and $\beta$\,$=$\,0.5 (Model4; dash-dot-dot-dot line) still does not provide an acceptable fit. Increasing $\beta$ even further, thereby extending the acceleration region, does not improve the fit, and also gives too narrow lines for the single-dish radio-line spectra.

Keeping Eq.~\ref{velfield}, with $\beta$ as a free parameter, and fitting it to a dynamical model of IRC\,+10216 performed by \citet{Ramstedt06b} a value of $\beta$\,$\approx$\,0.15 provides a good fit. A lower value for $\beta$ means a smaller acceleration region.  We note that \citet{Keady88} and \citet{Keady93} derive a more elaborate velocity field that also points to a small acceleration region. However, their results were not based upon any dynamical model. It turns out that a model with a $\beta$ smaller than 0.15 provides a worse fit.  The velocity of the wind is within 10\% of its terminal value at distances $>$\,5\,$\times$\,10$^{14}$\,cm. 

The only way to improve the fit to the observed ro-vibrational transitions is to increase the SiO fractional abundance in the acceleration region.
 If a model with a compact ($r_{\mathrm e}$\,$=$\,4.5\,$\times$\,10$^{14}$\,cm) high-fractional-abundance ($f_{\mathrm{J}}$\,$=$\,1.5\,$\times$\,10$^{-6}$) component is used a much better fit   to the data is obtained (Model 5, which includes an extended low abundance component, see below, that does not contribute to the ro-vibrational lines; solid line in Fig.~\ref{rovib}). 
 However, the region of high abundance can not be extended all the way down to the photosphere. The velocity field inside the dust condensation radius ($r_0$\,$=$\,1.7\,$\times$\,10$^{14}$\,cm) is assumed to be 2\,km\,s$^{-1}$ in accordance with \citet{Keady88}.  We emphasize that this is just a test and that the structure of the velocity field close to the photosphere certainly is more complex as shown by \citet{Winters00}.
 A much lower value of $\approx$\,3.0\,$\times$\,10$^{-8}$ is required in this region otherwise too much absorption will result in the ro-vibrational lines at velocities $\lesssim$\,2\,km\,s$^{-1}$. The adopted velocity field, resulting H$_2$ density structure from the continuity equation, and the derived SiO fractional abundance structure are shown in Fig.~\ref{structure}.
  
An extended low abundance component is still needed in order to model the single-dish radio data and SMA data. Using $f_0$\,$=$\,1.7\,$\times$\,10$^{-7}$ and $r_{\mathrm{e}}$\,$=$\,2.4\,$\times$\,10$^{16}$\,cm in addition to the compact component gives a good overall fit to the SMA data (Model 5; solid line), except that the model provides to much self-absorption at the extreme blue-shifted emission. The models are summarized in Table~\ref{models}. 
The two-component model provides the best fit to the observed size-velocity relation and the flux density profile as shown in Fig.~\ref{sma_size}.
Fig.~\ref{single} shows that the two-component model  also fits the available single-dish spectra well, with a reduced $\chi^2$ of 0.8 (only the total integrated intensity of the line was used for this estimate). Intensities for three additional lines other than the ones shown in Fig.~\ref{single} were used in the analysis \citep[see][]{Schoeier06a}. Also, the fit to the central pixel spectrum in the CLEANed image is excellent, as shown in Fig.~\ref{sma_spectrum}.

\section{Discussion}
\label{sect_discussion}
LTE stellar atmosphere models predict  that the SiO fractional abundance in carbon stars is relatively low, typically $\sim$\,5\,$\times$\,10$^{-8}$ \citep[see reviews by][ and references therein]{Glassgold99, Millar03} about three orders of magnitude lower than in M-type AGB stars.
The high SiO fractional abundances, derived in Sect.~\ref{sect_model}, of 1.5\,$\times$\,10$^{-6}$ in the inner wind ($\approx$\,3$-$8 stellar radii) of IRC\,+10216 can generally not be explained by LTE chemistry.

Departure from LTE could be caused by the variable nature of AGB stars that induces shocks propagating through the photosphere thereby affecting its chemistry. Models of shocked carbon-rich stellar atmospheres indicate that the SiO fractional abundance can be significantly increased by the passage of periodic shocks \citep{Willacy98, Helling01, Cherchneff06} . There is also a strong dependence on the shock strength and possibly this mechanism can explain the observed fractional abundances, and their large spread, of SiO in carbon stars \citep{Schoeier06a}. The lack of including absorption of molecules onto dust grains is a shortcoming of current generation chemical models  and will be required in a full model describing the chemistry of AGB stars. In addition, grain surfaces could act as catalysts for chemical reactions. 

The model that best explains all available observational constraints contains a compact high abundance component in addition to a more extended low-abundance component as illustrated in Sect.~\ref{sect_model}. This is the same behaviour as observed by \citet{Schoeier04b} for the low-mass-loss-rate M-type stars {R~Dor} and {L$^2$ Pup}.  Sch\"oier et al.\  introduced a compact component ($r_{\mathrm{e}}$\,$\approx$\,1\,$\times$\,10$^{15}$\,cm) with a high fractional abundance ($f_0$\,$\approx$\,4\,$\times$\,10$^{-5}$) to explain the small-scale emission. The motivation for this is that SiO is expected to effectively freeze out onto dust grains in the inner envelope. For a high-mass-loss-rate object such as IRC\,+10216 the freeze out is expected to be very efficient and simple condensation theory (see \citet{Delgado03b} for details) suggests that the SiO fractional abundance should be significantly affected within $r_{\mathrm{e}}$\,$\approx$\,5\,$-$\,10\,$\times$\,10$^{14}$\,cm. 

We also find evidence for a sharp decline of the SiO fractional abundance close to the photosphere ($\lesssim$\,3 stellar radii) from a value of $\approx$\,1.5\,$\times$\,10$^{-6}$ down to $\approx$\,3\,$\times$\,10$^{-8}$. This lower value is consistent with predictions from LTE stellar atmosphere models and could help explain why there are no detections of SiO maser emission in IRC\,+10216, or any other carbon star, as opposed to M-type AGB stars \citep[e.g.,][]{Lepine78}.



\section{Conclusions}
Interferometric sub-millimeter SiO $J$\,$=$\,5\,$\rightarrow$\,4 line observations performed at the SMA are presented for the carbon star {IRC+10216}. 
Based on a detailed excitation analysis and additional IR observations, a fractional abundance of SiO in the inner wind of IRC\,+10216 of more than an order of magnitude higher than predicted by thermal equilibrium chemistry is derived.  A possible explanation for the high SiO fractional abundance found is a shock-induced chemistry. However,  the influence of dust grains, both as a source for depletion as well as production of SiO, needs to be further investigated.  The infrared ro-vibrational data also provide information of the wind dynamics close to the star. The velocity of the wind has reached within 10\% of its terminal value already at a distance of 5\,$\times$\,10$^{14}$\,cm.

In addition to the compact high abundance component a spatially more extended region with low SiO abundance is needed to explain the observations. We find it likely that the observed abundance distribution around IRC\,+10216 is  the result of SiO depletion in the wind at distances $\gtrsim$\,8 stellar radii.
Similar abundance gradients have been reported for two M-type AGB stars by \citet{Schoeier04b}. High spatial resolution interferometric observations for a larger sample of sources could help to solidify this claim.

Evidence of departure from a smooth wind is found in the observed visibilities indicative of density variations a factor 2 to 5 on time scales of about 200 years, consistent with the density modulations found from scattered light measurements \citep{Mauron99,Mauron00}. 
 
\begin{acknowledgements}
%
We wish to thank an anonymous referee for useful comments.
FLS and HO acknowledge financial support from the Swedish Research Council.
\end{acknowledgements}


\begin{thebibliography}{36}
\expandafter\ifx\csname natexlab\endcsname\relax\def\natexlab#1{#1}\fi

\bibitem[{{Cherchneff}(2006)}]{Cherchneff06}
{Cherchneff}, I. 2006, \aap, {in press}

\bibitem[{{Cherchneff} \& {Barker}(1992)}]{Cherchneff92}
{Cherchneff}, I. \& {Barker}, J.~R. 1992, \apj, 394, 703

\bibitem[{{Duari} {et~al.}(1999){Duari}, {Cherchneff}, \& {Willacy}}]{Duari99}
{Duari}, D., {Cherchneff}, I., \& {Willacy}, K. 1999, \aap, 341, L47

\bibitem[{{Glassgold}(1996)}]{Glassgold96}
{Glassgold}, A.~E. 1996, \araa, 34, 241

\bibitem[{{Glassgold}(1999)}]{Glassgold99}
{Glassgold}, A.~E. 1999, in IAU Symp. 191: Asymptotic Giant Branch Stars, 337

\bibitem[{{Gonz{\' a}lez Delgado} {et~al.}(2003){Gonz{\' a}lez Delgado},
  {Olofsson}, {Kerschbaum}, {Sch{\" o}ier}, {Lindqvist}, \&
  {Groenewegen}}]{Delgado03b}
{Gonz{\' a}lez Delgado}, D., {Olofsson}, H., {Kerschbaum}, F., {et~al.} 2003,
  \aap, 411, 123

\bibitem[{{Groenewegen} {et~al.}(1998){Groenewegen}, {van der Veen}, \&
  {Matthews}}]{Groenewegen98}
{Groenewegen}, M.~A.~T., {van der Veen}, W.~E.~C.~J., \& {Matthews}, H.~E.
  1998, \aap, 338, 491

\bibitem[{{Groenewegen} \& {Whitelock}(1996)}]{Groenewegen96}
{Groenewegen}, M.~A.~T. \& {Whitelock}, P.~A. 1996, \mnras, 281, 1347

\bibitem[{{Habing} {et~al.}(1994){Habing}, {Tignon}, \& {Tielens}}]{Habing94}
{Habing}, H.~J., {Tignon}, J., \& {Tielens}, A.~G.~G.~M. 1994, \aap, 286, 523

\bibitem[{{Hasegawa} {et~al.}(2006){Hasegawa}, {Kwok}, {Koning}, {Volk},
  {Justtanont}, {Olofsson}, {Sch\"oier}, {Winnberg}, {Nyman}, {Frisk},
  {Hjalmarson}, {Olberg}, \& {Sandqvist}}]{Hasegawa06}
{Hasegawa}, T., {Kwok}, S., {Koning}, N., {et~al.} 2006, \apj, 637, 791

\bibitem[{{Helling} \& {Winters}(2001)}]{Helling01}
{Helling}, C. \& {Winters}, J.~M. 2001, \aap, 366, 229

\bibitem[{{Herbig} \& {Zappala}(1970)}]{Herbig70}
{Herbig}, G.~H. \& {Zappala}, R.~R. 1970, \apjl, 162, L15

\bibitem[{{Ho} {et~al.}(2004){Ho}, {Moran}, \& {Lo}}]{Ho04}
{Ho}, P.~T.~P., {Moran}, J.~M., \& {Lo}, K.~Y. 2004, \apjl, 616, L1

\bibitem[{{Ivezi{\' c}} \& {Elitzur}(1997)}]{Ivezic97}
{Ivezi{\' c}}, {\v{Z}}. \& {Elitzur}, M. 1997, MNRAS, 287, 799

\bibitem[{{Keady} {et~al.}(1988){Keady}, {Hall}, \& {Ridgway}}]{Keady88}
{Keady}, J.~J., {Hall}, D.~N.~B., \& {Ridgway}, S.~T. 1988, \apj, 326, 832

\bibitem[{{Keady} \& {Ridgway}(1993)}]{Keady93}
{Keady}, J.~J. \& {Ridgway}, S.~T. 1993, \apj, 406, 199

\bibitem[{{Lepine} {et~al.}(1978){Lepine}, {Scalise}, \& {Le
  Squeren}}]{Lepine78}
{Lepine}, J.~R.~D., {Scalise}, E., \& {Le Squeren}, A.~M. 1978, \apj, 225, 869

\bibitem[{{Mauron} \& {Huggins}(1999)}]{Mauron99}
{Mauron}, N. \& {Huggins}, P.~J. 1999, \aap, 349, 203

\bibitem[{{Mauron} \& {Huggins}(2000)}]{Mauron00}
{Mauron}, N. \& {Huggins}, P.~J. 2000, \aap, 359, 707

\bibitem[{{Melnick} {et~al.}(2001){Melnick}, {Neufeld}, {Ford}, {Hollenbach},
  \& {Ashby}}]{Melnick01}
{Melnick}, G.~J., {Neufeld}, D.~A., {Ford}, K.~E.~S., {Hollenbach}, D.~J., \&
  {Ashby}, M.~L.~N. 2001, \nat, 412, 160

\bibitem[{{Millar}(2003)}]{Millar03}
{Millar}, J. 2003, in Asymptotic giant branch stars, ed. H.~J. {Habing} \&
  H.~{Olofsson} (Astronomy and astrophysics library, New York, Berlin:
  Springer), 247

\bibitem[{{Miller}(1970)}]{Miller70}
{Miller}, J.~S. 1970, \apjl, 161, L95

\bibitem[{{Monnier} {et~al.}(2000){Monnier}, {Danchi}, {Hale}, {Tuthill}, \&
  {Townes}}]{Monnier00}
{Monnier}, J.~D., {Danchi}, W.~C., {Hale}, D.~S., {Tuthill}, P.~G., \&
  {Townes}, C.~H. 2000, \apj, 543, 868

\bibitem[{{Olofsson}(2006)}]{Olofsson06}
{Olofsson}, H. 2006, Reviews of Modern Astronomy, 19, {in press}

\bibitem[{{Ramstedt} {et~al.}(2006){Ramstedt}, {Sch{\"o}ier}, \&
  {Olofsson}}]{Ramstedt06b}
{Ramstedt}, S., {Sch{\"o}ier}, F.~L., \& {Olofsson}, H. 2006, \aap, {accepted}

\bibitem[{{Sch{\" o}ier} \& {Olofsson}(2000)}]{Schoeier00}
{Sch{\" o}ier}, F.~L. \& {Olofsson}, H. 2000, \aap, 359, 586

\bibitem[{{Sch{\" o}ier} \& {Olofsson}(2001)}]{Schoeier01}
{Sch{\" o}ier}, F.~L. \& {Olofsson}, H. 2001, \aap, 368, 969

\bibitem[{{Sch{\" o}ier} {et~al.}(2004){Sch{\" o}ier}, {Olofsson}, {Wong},
  {Lindqvist}, \& {Kerschbaum}}]{Schoeier04b}
{Sch{\" o}ier}, F.~L., {Olofsson}, H., {Wong}, T., {Lindqvist}, M., \&
  {Kerschbaum}, F. 2004, \aap, 422, 651

\bibitem[{{Sch{\" o}ier} {et~al.}(2002){Sch{\" o}ier}, {Ryde}, \&
  {Olofsson}}]{Schoeier02b}
{Sch{\" o}ier}, F.~L., {Ryde}, N., \& {Olofsson}, H. 2002, \aap, 391, 577

\bibitem[{{Sch{\" o}ier} {et~al.}(2005){Sch{\" o}ier}, {van der Tak}, {van
  Dishoeck}, \& {Black}}]{Schoeier05a}
{Sch{\" o}ier}, F.~L., {van der Tak}, F.~F.~S., {van Dishoeck}, E.~F., \&
  {Black}, J.~H. 2005, \aap, 432, 369

\bibitem[{{Sch{\"o}ier} {et~al.}(2006){Sch{\"o}ier}, {Olofsson}, \&
  {Lundgren}}]{Schoeier06a}
{Sch{\"o}ier}, F.~L., {Olofsson}, H., \& {Lundgren}, A.~A. 2006, \aap,
  {in press (astro-ph/0604213)}

\bibitem[{{van Zadelhoff} {et~al.}(2002){van Zadelhoff}, {Dullemond}, {van der
  Tak}, {Yates}, {Doty}, {Ossenkopf}, {Hogerheijde}, {Juvela}, {Wiesemeyer}, \&
  {Sch{\" o}ier}}]{Zadelhoff02}
{van Zadelhoff}, G.-J., {Dullemond}, C.~P., {van der Tak}, F.~F.~S., {et~al.}
  2002, \aap, 395, 373

\bibitem[{{Willacy}(2004)}]{Willacy04}
{Willacy}, K. 2004, \apjl, 600, L87

\bibitem[{{Willacy} \& {Cherchneff}(1998)}]{Willacy98}
{Willacy}, K. \& {Cherchneff}, I. 1998, \aap, 330, 676

\bibitem[{{Winters} {et~al.}(1994){Winters}, {Dominik}, \&
  {Sedlmayr}}]{Winters94}
{Winters}, J.~M., {Dominik}, C., \& {Sedlmayr}, E. 1994, \aap, 288, 255

\bibitem[{{Winters} {et~al.}(2000){Winters}, {Keady}, {Gauger}, \&
  {Sada}}]{Winters00}
{Winters}, J.~M., {Keady}, J.~J., {Gauger}, A., \& {Sada}, P.~V. 2000, \aap,
  359, 651

\end{thebibliography}
\end{document}